\documentclass[12pt,aps,ams,amsfonts,nofootinbib]{revtex4}

\usepackage[dvips]{graphics}

\usepackage{graphicx}
\usepackage{amssymb}
\usepackage{subfigure}
\usepackage{stackrel}
\usepackage{enumitem}
\usepackage{amsmath}
\usepackage{amsfonts}
\usepackage{bm}
\usepackage{color}
\usepackage{verbatim} %for comments
\usepackage{tcolorbox}
\usepackage[normalem]{ulem}
\usepackage{hyperref}
\usepackage{comment}

\newlist{subquestion}{enumerate}{1}
\setlist[subquestion,1]{label=(\alph*)}

\newcommand{\ket}[1]{\ensuremath{\left|{#1}\right\rangle}}
\newcommand{\bra}[1]{\ensuremath{\left\langle{#1}\right |}}
\newcommand{\operdef}[2]{\ensuremath{|{#2}\rangle \langle{#1}|}}

\newcommand{\beq}{\begin{equation}}
\newcommand{\eeq}{\end{equation}}
\newcommand{\bse}{\begin{subequations}}
	\newcommand{\ese}{\end{subequations}}\newcommand{\bea}{\begin{eqnarray}}
\newcommand{\eea}{\end{eqnarray}}
\newcommand{\bit}{\begin{itemize}}
	\newcommand{\eit}{\end{itemize}}
\newcommand{\bpmatrix}{\begin{pmatrix}}
	\newcommand{\epmatrix}{\end{pmatrix}}

\newcommand{\be}{\begin{equation}}
\newcommand{\ee}{\end{equation}}
\newcommand{\ben}{\begin{eqnarray}}
\newcommand{\een}{\end{eqnarray}}

\begin{document}

\title{Generalized approach to quantify correlations in bipartite quantum systems}

\author{D. G. Bussandri$^{1,2}$ , A. P. Majtey$^{1,3}$, P. W. Lamberti$^{1,2}$, T. M. Os\'an$^{1,3}$}
\affiliation{$^1$Facultad de Matem\'atica, Astronom\'{\i}a, F\'{\i}sica y Computaci\'on, Universidad Nacional de C\'ordoba, Av. Medina Allende s/n, Ciudad Universitaria, X5000HUA C\'ordoba, Argentina}
\affiliation{$^2$Consejo Nacional de Investigaciones Cient\'{i}ficas y T\'ecnicas de la Rep\'ublica Argentina, Av. Rivadavia 1917, C1033AAJ, CABA, Argentina}
\affiliation{$^3$ Instituto de F\'isica Enrique Gaviola, Consejo Nacional de Investigaciones Cient\'{i}ficas y T\'ecnicas de la Rep\'ublica Argentina, Av. Medina Allende s/n, X5000HUA, C\'rdoba, Argentina}

%\date{\today}

\begin{abstract}
In this work we developed a general approach to the problem of detecting and quantifying different types of correlations in bipartite quantum systems. Our method is based on the use of distances between quantum states and processes. We rely upon the premise that total correlations can be separated into classical and quantum contributions due to their different nature. In addition, 
according to recently discussed criteria,  we determined the requirements to be satisfied by distances in order to generate correlation measures physically well--behaved. The proposed measures allow us to quantify quantum, classical and total correlations. Besides the well-known case of relative entropy we introduce some additional examples of distances which can be used to build \textit{bona fide} quantifiers of correlations.
\end{abstract}

%\pacs{03.67.Mn, 03.65.Ud}% PACS, the Physics and Astronomy

\maketitle                           %display desi d
\section{Introduction\label{sec:intro}}

Quantum information processing and quantum computing  are relatively new subjects of research concerned with the use of quantum resources
to perform tasks of information processing which are either not feasible to be implemented classically or can be performed with classical devices in a much less efficient way. Since the inception of these research fields, the central point seems to be able to identify which features in the quantum realm are responsible for the so-called \textit{quantum advantage}. Among the possible quantum features, entanglement is regarded as the main resource for practical use in quantum information processing \cite{vedral_2014,yu_2007,yu_2009,plenio_2007,brunner_2014,scirep5_2015,scirep5_2015b,prl104_2010,aolita_2015,lofranco_2013,lofranco_2013b,lofranco_2014,lofranco_2016,lofranco_2017}, particularly, for quantum  protocols in quantum networks (composite systems). However, at present it is widely recognized that this is not the unique asset which can be used for quantum protocols in order to outperform their classical counterparts. Several results allow us to conclude that, in some quantum tasks, correlations of a quantum nature different from entanglement are responsible for the processing improvements \cite{Knill98,LCNV,Braun99,Meyer00,Datta05,Datta07,Datta08,Lanyon08}. In fact, different works on quantum correlations have been devoted to analyze their resilience to noise under certain conditions \cite{silva_2016,aaronson_2013,aaronson_2013b,haikka_2013,cianciaruso_2015}. Also, it is worth mentioning that quantum nonlocality may also arise even in the absence of entanglement \cite{Bennet1999}.   \par
Usually, because of the distinct nature of quantum and classical correlations \cite{Grois2005,HV01,Horod2002,Horod2008}, the procedures used to detect the presence of correlations rely upon the reasonable hypothesis that total correlations $\mathcal{T}$ contained in a bipartite quantum state $\rho$ 
can be quantified as follows
\begin{equation}
\mathcal{T}(\rho)=\mathcal{Q} (\rho)+\mathcal{J}(\rho)\label{eq:gentotcorr},
\end{equation}
\noindent where $\mathcal{Q}$ and $\mathcal{J}$ represent measures of quantum and classical correlations, respectively. It is important to realize that, within this framework, Eq. \eqref{eq:gentotcorr} makes sense if all three quantities are measured in the same units.  In particular, a widely accepted information-theoretic measure of total correlations contained in a bipartite quantum state $\rho$ is (von Neumann) Quantum Mutual Information $\mathcal{I}(\rho)$ defined as:
\begin{align}
\mathcal{I}(\rho) \doteq S(\rho_A) + S(\rho_B) -
S(\rho).
\label{eq:QMI}
\end{align}
In Eq. \eqref{eq:QMI}, $\rho$ stands for a general bipartite quantum state, 
$\rho_A=Tr_B\left[\rho_{AB} \right]$ and $\rho_B=Tr_A\left[\rho_{AB} \right]$ represent the corresponding reduced (marginal) states, and $S(\rho)$ represents the von Neumann entropy given by
\begin{align}
S(\rho) = -\mbox{Tr} \left[ \rho \log_2 \rho\right] .
\label{def8}
\end{align}
The quantity $\mathcal{I}(\rho)$ describes the correlations between the whole subsystems rather than a  correlation between just two observables.\par
Classical correlations contained in a quantum state $\rho$ of a bipartite quantum system can be quantified, for example, by means of the measure $\mathcal{J}_S(\rho)$ proposed by Henderson and Vedral in ref. \cite{HV01}. Thus, whenever total correlations are measured with mutual information $\mathcal{I}(\rho)$ and classical correlations are measured according to the aforementioned quantity $\mathcal{J}_S(\rho)$ (cf. Sec. \ref{sec:Jvedral}), the measure of quantum correlations that results is the widely known definition of \textit{quantum discord} (QD) \cite{OZ02,HV01,Luo08b}, i.e.,
\begin{align}
\mathcal{D} (\rho)=\mathcal{I}(\rho)-\mathcal{J}_S(\rho)\label{eq:QDdef}. 
\end{align}
As we will see in section~\ref{sec:Jvedral}, relation~\eqref{eq:gentotcorr} lies at the core of the theoretical framework used in the present paper. However, it should be noticed that Eq. \eqref{eq:gentotcorr} may or may not be satisfied, depending on the approach used to quantify the different type of correlations present in the system under study. For example, when a geometric scheme is used, such as the one introduced in~\cite{Modi2010}, equation~\eqref{eq:gentotcorr} is not satisfied. Though, it may be mentioned that a geometric approach to determine correlations should be used with caution since in general it presents some peculiar behaviors, such as those discussed in references~\cite{{Piani2012,Tufarelli2013,Paula2014}}, for example.
It can be shown that the essence of QD, as defined by Eq.~\eqref {eq:QDdef}, is to quantify the discrepancy between the quantum versions of two classically equivalent expressions for mutual information~\cite{OZ02,HV01}. However, although from a conceptual point of view QD is of relevance in assessing possible non-classical resources for information processing, for practical use it presents some drawbacks. For example, at this moment there is no straightforward criterion to verify the presence of nonzero discord in a given \textit{arbitrary} bipartite quantum state  (i.e., an arbitrary bipartite state belonging to the product of two Hilbert spaces of arbitrary dimensions). Besides, as the evaluation of QD involves an optimization procedure, analytical results are only known in some particular cases~\cite{Lang10,Cen11,Adesso10,Ali10,Shi11,Chen11,Lu11,Giro12,Li11,Luo08b}. Furthermore, in general, calculation of quantum discord is NP-complete since the optimization procedure needs to be done by means of a sweep over a complete set of measurements performed over one of the subsystems \cite{Huang14}. Additionally, quantum discord can be created starting from a state with \textit{zero} discord and then  performing Local Operations \cite{DVB2010,Gessner2012}. It is hard to accept that local operations alone do increase the amount of correlations between two subsystems. Instead, it is expected that correlations contained in the resulting state are lower than or equal to the correlations of the classical state. As this last statement is not reflected in the behaviour of quantum discord, it turns out to be questionable to affirm that (all) discordant states can be thought of as resources useful for quantum information processing tasks.
Therefore, a central problem in quantum physics is to be able to characterize and quantify correlations in multipartite states. Further research on correlations is important not only from a practical point of view but also from a conceptual point of view as they can provide additional insights about the underlying physics behind correlations present  in quantum systems. Detecting the existence of non-classical correlations in a given system is the central goal for a measure of correlations regarding the possibility of classifying a given system as a genuinely quantum one. To this end, it may result of interest to devise proper means to detect and quantify the presence of such correlations.
Measures of different type of correlations in quantum systems can be motivated by different notions of classicality or operational means to quantify nonclassicality. 
Thus, measures of quantum correlations may in general differ, both formally and conceptually.
Regarding quantum discord, for example,  it can be seen that it does not coincide in general with entanglement or measurement induced disturbance \cite{Modi2010,Modi2012,Luo08c}. Moreover, a direct comparison between these two notions may be meaningless \cite{Modi2010,Modi2011}. As a result, general approaches and theoretical frameworks seem to be central to the study and understanding of different notions of quantumness.\par
In previous works, some authors thoroughly discussed a set of properties that \textit{bonafide} measures of correlations should satisfy \cite{Brodutch12,ABC16,cianciaruso_2015}. In this work we introduce a scheme to measure correlations in bipartite quantum systems in the following way: Starting from the (plausible) definition of a measure of total correlations $\mathcal{T}(\rho)$  and a measure of classical correlations, we build a measure $\mathcal{Q}(\rho)$ of quantum correlations based on the difference between the two former. Then, based on refs. \cite{Brodutch12} and \cite{ABC16}, we show that the measure $\mathcal{Q}(\rho)$ satisfies a set of desirable properties for a suitable measure of quantum correlations. In this way, our proposal for measures of different type of correlations in a bipartite quantum system turns out to be consistent.\par
This paper is organized as follows. In Sec. \ref{sec:theory}, we outline the basic theoretical background directly related to our work. In Sec. \ref{sec:main}, we develop our main results, i.e., we introduce a general approach to define measures of different type of correlations using the concept of distinguishability between quantum states. Finally, some conclusions are drawn in Sec. \ref{sec:conclusions}.
\section{Theoretical framework\label{sec:theory}}
%%%%%%%%%%%%%%%%%%%%%%%%%%%%%
\subsection{Distance measures in the quantum realm \label{sec:dprop}}
From a conceptual point  of view, a \textit{metric} $d(.||.)$ on a set $\chi$ is a functional $d\!:\!\chi\!\times\!\chi\!\rightarrow\!\mathbb{R}_{\geq 0}$ such that for
every $x,y,z \in \chi$ the following properties are satisfied:
\begin{enumerate}[label=\alph*)]
	\item \label{M1} {\em Non-negativity}: For any $x,y \in \chi$, $d(x||y) \geq 0$ \label{TP2}
	\item \label{M2} {\em Identity of indiscernibles}: $d(x||y) = 0$ if and only if $x=y$ \label{TP1}
	\item \label{M3} {\em Symmetry}: For any $x,y \in \chi$, $d(x||y) = d(y||x)$ 
	\item \label{M4} {\em Triangle inequality} For any $x,y,z \in \chi$, $d(x||y) \leq d(x||z) + d(z||y)$ 
\end{enumerate}
In the context of probability theory, a functional $d$  satisfying properties \ref{M1} and \ref{M2} is called a {\em divergence}. If, in addition, $d$ satisfies property \ref{M3}, then $d$ is called a {\em distance}.\par
In quantum physics, the divergences and distances are defined on the set of quantum states represented by density operators acting on a Hilbert space $\mathcal{H}$. Density operators representing the states of a given system are elements belonging to $\mathcal{B}^+_1(\mathcal{H})$, i.e., the set of bounded, positive-semidefinite operators on $\mathcal{H}$, with unit trace. In particular, unit vectors in $\mathcal{H}$ correspond to the extremal elements of $\mathcal{B}^+_1(\mathcal{H})$ ($\rho \in \mathcal{B}^+_1(\mathcal{H})$ is extremal if and only if it is idempotent, i.e., $\rho^2 = \rho$). Any pure state corresponds to a unit vector $\ket{\varphi} \in \mathcal{H}$ and can be represented by a density matrix of the form $\rho = \operdef{\varphi}{\varphi}$.\par
One of the most striking features of quantum mechanics is that, in general, two arbitrary quantum states cannot be distinguished with certainty. Only orthogonal states can be discriminated unambiguously. Therefore, in order to determine how close two quantum states are to each other, a variety of distance measures have been developed, as for example, {\em trace distance}, {\em Fidelity}, {\em Bures distance}, {\em Hilbert-Schmidt distance}, {\em Hellinger distance} and {\em Quantum Jensen-Shannon divergence}, just to name a few \cite{Nielsen&Chuang,BZ06,Uhlm1976,Jozsa1994,Bures1969,Dodo2000,Luo2004,Majtey2005}.\par
From physical grounds, the following properties are also usually required for a distance between quantum states to be well-behaved \cite{Nielsen&Chuang}:
\begin{enumerate}[label=\alph*),resume]
	\item $d(\cdot || \cdot)$ must be invariant under unitary transformations, i.e. \label{AP1} \label{TP3}
	\begin{align}
	d(U\rho U^\dagger || U\sigma U^\dagger)=d( \rho || \sigma ),
	\end{align} 
	being $U$ some unitary operator;
	\item \label{PRTC} $d(\cdot||\cdot)$ must be non-increasing under the action of a trace-preserving quantum operation, 
	\begin{align}
	d(\mathcal{E}(\rho)||\mathcal{E}(\sigma))\leq d(\rho||\sigma),
	\end{align}
	\item \label{PRCC} $d(\cdot||\cdot)$ must be convex in one of its inputs,
	\begin{align}
	d( \  \sum_i p_i\rho_i \ || \ \sigma \ )\leq \sum_i p_i d(\rho_i||\sigma),
	\end{align}
	\noindent being $\{p_i\}$ a probability distribution and $\{\rho_i\}$ elements of $\mathcal{B}_1^+(H)$;
\end{enumerate}
In addition, as we shall discuss in Sec.\ref{sec:GenMeasurQcorr}, the following property shall be required for well-behaved measures of quantum correlations:
\begin{enumerate}[label=\alph*),resume]
	\item \label{PRQC} $d(\cdot||\cdot)$ must satisfy
	\begin{align}
	&d( \ \sum_i p_i \left|i\right>\left<i\right| \otimes \rho_i \ || \ \sum_j p_j \left|j\right>\left<j\right| \otimes \rho ) = \sum_k p_k d(\rho_i || \rho),
	\end{align}
	being $\{\left|i\right>\}$ an orthonormal basis of $\mathcal{H}_A$,  $\{p_i\}$ a probability distribution and $\{\rho_i\}$ elements of $\mathcal{B}_1^+(H)$ such that $\sum_i  p_i \rho_i = \rho$.
\end{enumerate}
\subsection{Correlations in quantum systems}\label{sec:Jvedral}
Let us consider a bipartite system ($A+B$) in an arbitrary quantum state $\rho$ belonging to $ \mathcal{B}^+_1(\mathcal{H}_{AB})$, being $\mathcal{H}_{AB} =\mathcal{H}_A \otimes\mathcal{H}_B$ and $\mathcal{H}_A$ and $\mathcal{H}_B$ the Hilbert spaces associated with subsystems $A$ and $B$, respectively.\par
If the system is prepared in a pure state $|\psi\rangle_{AB} \in \mathcal{H}_{AB}$, (i.e., $\rho = \operdef{\psi_{AB}}{\psi_{AB}}$) we have essentially two relevant physical scenarios. In the first one,
the subsystems are independent (i.e., there are no correlations of any type between both subsystems). In this case, the state can be described as a tensor product, i.e.,  $|\psi\rangle_{AB}=|\phi\rangle_A\otimes|\chi\rangle_B$. In the second scenario, the global state cannot be written as a tensor product, i.e., $|\psi\rangle_{AB}\neq|\phi\rangle_A\otimes|\chi\rangle_B$. In this case, the state is said to be entangled. Entanglement is the unique form of quantum correlation that might take place in the case of bipartite pure states. Moreover, entanglement, non locality, and QCs are synonymous in this particular case.\par
In contrast, for a system prepared in a mixed state described by a density matrix $\rho$, other type of correlations can be involved. In this case we are in the presence of a subtler and richer situation than the case of pure states. Indeed, unentangled mixed states can exhibit correlations of a quantum nature \cite{Knill98,Lanyon08}.\par
The set of \textit{separable} (unentangled) states $\mathcal{S}_{A,B}$ is composed by density operators $\rho \in \mathcal{B}_1^+(\mathcal{H}_{AB})$ such that
\begin{equation} \label{separables}
\mathcal{S}_{A,B} \doteq \{ \rho \in \mathcal{B}_1^+(\mathcal{H}_{AB}) \ / \ \rho=\sum_i \ p_i \ \rho_A^i \otimes \rho_B^i \}
\end{equation}
where $\rho^i_{A(B)}$ is associated with the quantum state of subsystem $A$ ($B$) and $\{p_i\}$ are non-negative numbers such that $\sum_i p_i =1$.
The set of product states (which is included in the set of separable states) is defined as follows
\begin{equation}\label{producto}
\mathcal{P}_{A,B} \doteq \{ \rho \in \mathcal{B}_1^+(\mathcal{H}_{AB})\ / \ \rho=\rho_A \otimes \rho_B \}.
\end{equation}
A state $\rho$ belonging to the set $\mathcal{P}_{A,B}$ is called \textit{uncorrelated} in the sense that it does not contain correlations of any type whatsoever. Then, a given quantum state $\rho$ can be a separable or a product state if it can be reduced to the form established in equations \eqref{separables} or \eqref{producto}, respectively. \par
Any state $\rho$ that can not be written in the form given by Eq. \eqref{separables} is referred to as non-separable. It is important to realize that only non-separable states have quantum correlations analogous to that of an entangled pure state. However, non-classical correlations may be present in some subset of separable states.
Regarding a bipartite system where subsystems $A$ and $B$ are classically correlated, it can be shown that the global state $\rho$ belong to the following set:
\begin{equation}
\mathcal{C}_{A,B} \doteq \{ \rho  \in \mathcal{B}_1^+(\mathcal{H}_{AB}) \ / \ \rho=\sum_{i,j} \ p_{i,j} \ \left| i \right> \left< i \right|_A  \otimes  \left| j \right> \left< j \right|_B \}
\end{equation}
\noindent being $\{\left| i \right>_A\}$ and  $\{\left| j \right>_B\}$ orthonormal bases of the Hilbert spaces $\mathcal{H}_A$ and $\mathcal{H}_B$, respectively, and $\{p_{ij}\}$ a joint probability distribution. Furthermore, the set of (asymmetric) classical-quantum states is defined as follows,
\begin{equation}\label{C_A}
\mathcal{C}_{A} := \{ \rho   \in \mathcal{B}_1^+(\mathcal{H}_{AB})\ / \ \rho=\sum_{i} \ p_{i} \ \left| i \right> \left< i \right|_A  \otimes  \rho_B^i \}
\end{equation}
\noindent where $\{p_{i}\}$ is a probability distribution and $\{ \rho_B^i \}$ any set of quantum states of the subsystem $B$. The set  $\mathcal{C}_B$ of quantum-classical states can be analogously defined.\par
Total correlations given by Eq. \eqref{eq:QMI} can be recast in terms of von Neumann relative entropy as follows:
\begin{align}
I(\rho) = S(\rho || \rho_A\otimes\rho_B), \label{eq:QMI2}
\end{align}
\noindent where relative entropy is given by
\begin{align}
S(\rho||\sigma) = \mbox{Tr} \left[ \rho (\ln \rho - \ln \sigma)\right] .
\label{QRelEnt}
\end{align}
Thus, Eq.\eqref{eq:QMI2} can be interpreted as the distance between the state $\rho$ and the uncorrelated state $\rho_A\otimes\rho_B$. In this case, the distance is quantified by means of quantum relative entropy.\par
According to ref. \cite{HV01}, the measure of classical correlations $\mathcal{J}_S(\rho)$ involved in Eq. \eqref{eq:QDdef} is defined as
\begin{align}
\mathcal{J}_S\left(\rho\right)=S(\rho_B)-\min_{M_j} \sum_j \ p'_j \ S (\rho_{B|j}^\mathcal{M} ), \label{Vedralc}
\end{align}
\noindent being $\mathcal{M}=\left\{M_j\right\}_{j=1}^{m}$ ($m\in\mathbb {N}$), a von Neumann measurement on subsystem $A$ (i.e., a family of orthonormal projectors on $\mathcal{H}_A$) and being
\begin{align}
\rho_{B|j} ^\mathcal{M}=\textrm{Tr}_A\left[ M_j \otimes \mathbb{I} \rho \right]/p'_j; \label{rhoBJ} \\
p'_j=\textrm{Tr}\left[ M_j \otimes \mathbb{I} \rho \right], \label{PprimaJ}
\end{align}
\noindent the states of the subsystem $B$ and the conditional probabilities of getting the state $\rho_{B|j} ^\mathcal{M}$ after the measurement $\mathcal{M}$, respectively. From its definition (cf. Eq. \eqref{Vedralc}), it can be seen that the measure $\mathcal{J}_S\left(\rho\right)$ it is not symmetric under the exchange of subsystems $A$ and $B$. Therefore,  in some sense there exists a \textit{directionality} over $\mathcal{J}_S\left(\rho\right)$ and, as a consequence over the quantity $\mathcal{D}(\rho)$. States given by Eq. \eqref{rhoBJ} are commonly referred to as \textit{conditional states}.\par
From Eq. \eqref{eq:gentotcorr} we can define a {\it directional} measure of quantum correlations in the form
\begin{equation}
\mathcal{Q} (\rho) \doteq \mathcal{T}(\rho) - \mathcal{J}(\rho)\label{eq:genQCorr}.
\end{equation}
From last equation, it can be easily seen that QD is a particular case of a measure of quantum correlations when we choose $\mathcal{T}(\rho) = \mathcal{I}(\rho)$ (cf. eqs. \eqref{eq:QMI} and \eqref{eq:QMI2}) and $\mathcal{J}(\rho)=\mathcal{J}_S(\rho)$ (cf. Eq. \eqref{Vedralc}). Bearing in mind \eqref{eq:genQCorr} we will introduce in Sec.\ref{sec:main} a general approach to define measures of different types of correlations based upon the concept of distinguishability between states in the quantum realm.\\
We would like to emphasize that $\mathcal{J}\left(\rho\right)$ is not a symmetric quantity under the exchange of $A$ for $B$. Therefore, $\mathcal{J}\left(\rho\right)$ is in fact a measure of classical-quantum correlations.
\subsection{Properties of correlation measures}\label{prop measures}
The set of criteria to be satisfied for valid quantifiers of correlations in quantum systems is still a subject of active research.  However, following refs. \cite{Brodutch12} and \cite{ABC16} we shall require the following set of \textit{necessary} properties for any well-behaved measure of correlations:
\begin{enumerate}
	\item Product states have no correlations, then: $\mathcal{Q} (\rho_A \otimes \rho_B)  = \mathcal{T}(\rho_A \otimes \rho_B)= \mathcal{J}(\rho_A \otimes \rho_B) = 0$;  \label{cond 1}
	\item All correlation measures should be invariant under local unitary operations; \label{cond 2}
	\item All correlation measures should be non-negative, i.e.,  $\mathcal{Q} (\rho) \geq 0$, $\mathcal{T}(\rho) \geq 0$ and $\mathcal{J}(\rho) \geq 0$; \label{cond 3}
	\item Total correlations measures should be non-increasing under local operations; \label{cond 4}
	\item Classical states have no quantum correlations. \label{cond 5}
\end{enumerate}
According to ref. \cite{Brodutch12}, any measure which does not meet these necessary criteria is not a  valid measure of correlations.\par	
\section{Generalized correlation measures in bipartite quantum systems\label{sec:main}}
In this section we propose measures of total, classical and quantum correlations. Also, taking into account the criteria previously introduced (cf. Sec.\ref{sec:theory}), we analyze the properties to be satisfied by a distance in order to build proper correlation measures.
\subsection{Generalized measures of total correlations}
Equation \eqref{eq:QMI2} provides the fundamental basis to define measures of total correlations by means of a distance between quantum states. Thus, given an arbitrary distance $d(\cdot||\cdot)$ we propose a measure of total correlations as follows:
\begin{equation}
\label{Info con d}
\mathcal{T}_d\left( \rho \right) = d\left(\rho||\rho_A \otimes \rho_B \right).
\end{equation}
The relation \eqref{Info con d} provides a whole new family of quantifiers which, in principle, might be used as tools for analyzing different possible scenarios of physical relevance.\par
Now, in order to prove that the measure $ \mathcal{T}_d(\rho) $ quantifies total correlations, we shall study how the properties imposed on $d(\cdot||\cdot)$ are related to the properties to be fulfilled by $\mathcal{T}_d(\rho)$. Thus, we establish the following propositions:
\begin{description}[font=\bfseries,align=left]
	\item[Proposition I] If $d(\cdot||\cdot)$ satisfies \ref{TP2}, \ref{TP1} and \ref{TP3} (cf. Sec.\ref{sec:dprop}) then $\mathcal{T}_d$ fulfills the conditions \ref{cond 1}-\ref{cond 3} (cf. Sec.\ref{prop measures}).
	\item[Proposition II] If, in addition, $d(\cdot||\cdot)$ also satisfies the 
	\ref{PRTC} (cf. Sec.\ref{sec:dprop}) then $\mathcal{T}_d$ fulfills the property \ref{cond 4} (cf. Sec.\ref{prop measures}).
\end{description}
The proofs of both propositions are straightforward.
\subsection{Generalized measures of classical-quantum correlations}
In this section our aim is to introduce classical-quantum correlations measures for a bipartite arbitrary state.  The basic idea is to apply a measurement $\mathcal{M}$ on subsystem $A$ and see afterwards how this action \textit{conditions} the state of subsystem $B$. Then, we shall quantify how different is the state $\rho_B$ from the states $\rho_{B|j}^\mathcal{M}$ arising after performing the local measurement on  $A$, by means of the distance $d(\rho_{B|j}^\mathcal{M} \ || \ \rho_B)$.
Thus, a measure $\mathcal{J}_d\left( \rho \right)$ of correlations between the two subsystems shall be the maximum of the weighted average of the above $m\in\mathbb{N}$ quantities $d(\rho_{B|j}^\mathcal{M} \ || \ \rho_B)$, i.e.,
\begin{align}
\label{cc}
\mathcal{J}_d\left( \rho \right) &:= \max_{\{\mathcal{M}\}} \ \mathcal{J}^{\mathcal{M}}_d \left( \rho \right),
\end{align}
being $\mathcal{J}^{\mathcal{M}}_d \left( \rho \right)$ defined as follows
\begin{align}
\label{ccm}\mathcal{J}^{\mathcal{M}}_d \left( \rho \right) &:=\sum_{j=1}^m p'_{j} \ d\left( \rho_{B|j}^\mathcal{M} \ || \ \rho_B \right).
\end{align}
Looking at equations \eqref{cc} and \eqref{ccm}, we notice that expression \eqref{Vedralc} is recovered whenever the distance $d(\cdot||\cdot)$ is replaced with the relative entropy. \par
Now, we shall study the necessary properties to be fulfilled by  $d(\cdot||\cdot)$ in order for $\mathcal{J}_d$ to meet the criteria for classical-quantum correlation measures. With this aim, we establish the following proposition. 
\begin{description}[font=\bfseries,align=left]
	\item[Proposition III] If $d(\cdot||\cdot)$ satisfies properties \ref{TP2}, \ref{TP1} and \ref{TP3} (cf. Sec. \ref{sec:dprop}), then $\mathcal{J}_d$ fulfills the conditions \ref{cond 1}-\ref{cond 3} (cf. Sec. \ref{prop measures}). 
\end{description}
\textbf{Proof:} Let us suppose that the system is in a product state $\sigma=\sigma_A\otimes\sigma_B$. In this case, it can be seen that $\sigma_{B|j}^\mathcal{M}=\sigma_B$ for any measurement $\mathcal{M}$. Thus, if the distance $d$ satisfy \ref{TP1}) (cf. Sec. \ref{sec:dprop}) it follows that $d\left( \sigma_{B|j}^\mathcal{M} \ || \ \sigma_B \right)=0$. Therefore, we have shown that $\mathcal{J}_d\left( \sigma \right)=0$, and the requirement \ref{cond 1} is fulfilled.\par
Next, we shall prove that requirement \ref{cond 2} is satisfied by $\mathcal{J}_d\left( \rho \right)$ whenever the distance $d$ satisfies property \ref{TP3}, i.e., it remains invariant under local unitary transformations.\par
Indeed, let us consider $\sigma=U \rho U^\dagger$ with $U=U_A \otimes U_B$. Then, $\sigma_{B|j}^\mathcal{M}=U_B \rho_{B|j}^\mathcal{M'} U_B^\dagger$ and $\sigma_B=U_B\rho_B U^\dagger_B$, being $\mathcal{M}'=\{U_A M_j U_A^\dagger \}$. Thus, it follows that
\begin{align}
d\left( \sigma_{B|j}^\mathcal{M} \ || \ \sigma_B \right)=d\left( U_B \rho_{B|j}^\mathcal{M'} U_B^\dagger \ || \ U_B \rho_B U_B^\dagger\right)=d(\rho_{B|j}^\mathcal{M'} \ || \ \rho_B)
\end{align}
Furthermore, it can be seen that the probability distribution $p'_j$ for the state $\sigma$ is $p'_j=\textrm{Tr}[M_j \otimes \mathbb{1}  \sigma]=\textrm{Tr}[M'_j \otimes \mathbb{1} \rho]$. Therefore, 
\begin{align}
\mathcal{J}_d^\mathcal{M}(\sigma)=\mathcal{J}_d^\mathcal{M'}(\rho)
\end{align}
Due to $\mathcal{M}'$ is composed by each one of the elements of $M$ transformed upon the action of the unitary operation $U_A$ on party $A$, we have:
\begin{align}
\mathcal{J}_d\left( \sigma \right)= \max_{\{\mathcal{M}\}} \ \mathcal{J}^{\mathcal{M}}_d \left( \sigma \right)=\max_{\{\mathcal{M}\}} \ \mathcal{J}^{\mathcal{M}'}_d \left( \rho \right)=\max_{\{\mathcal{M}\}} \ \mathcal{J}^{\mathcal{M}}_d \left( \rho \right)=\mathcal{J}_d\left( \rho \right)
\end{align}
Thus, we have shown that property \ref{cond 2} is satisfied by $\mathcal{J}_d\left( \rho \right)$.\par
Finally, the positivity of $\mathcal{J}_d\left( \rho \right)$ (cf. Sec.\ref{prop measures}, property \ref{cond 3} follows from equations \eqref{cc} and \eqref{ccm} and the positivity of $d$ (cf. \ref{TP2}, Sec.\ref{sec:dprop})$\blacksquare$.
 
\vspace{1cm}

In order for $\mathcal{J}_d$ to be a measure of classical-quantum correlations, let us consider the following classical-quantum state $\rho_c=\sum_{i}  p_{i}  \left| i \right> \left< i \right|_A  \otimes  \rho_B^i$. Then, the following condition needs to be satisfied,
\begin{align}\label{extra property}
\mathcal{J}_d(\rho_c)=\sum_i p_i \ d\left(\rho_B^i \ || \ \rho_B\right).
\end{align}
This last equation follows from the fact that for the state $\rho_c $ the best measurement we can perform on the subsystem $A$ to obtain information about the subsystem $B$ is a projective measurement on the states $\{\left|i\right>\}$ \cite{HV01}, i.e., $\mathcal{M}=\{\left|i\right>\left<i\right|\}$ maximizes \eqref{ccm}, therefore Eq. \eqref{cc} should match Eq. \eqref{extra property} for $\rho_c$.\par
Next, we shall study the properties to be satisfied by $d(\cdot||\cdot)$ in order for $\mathcal{J}_d$ to fulfill equation \eqref{extra property}. Thus, we establish the following proposition:
\begin{description}[font=\bfseries,align=left]
	\item[Proposition IV] If $d(\cdot||\cdot)$ is convex in one of its inputs, (cf. property \ref{PRCC}, Sec. \ref{sec:dprop}), then $\mathcal{J}_d(\rho_c)$ shall satisfy Eq. \eqref{extra property}.
\end{description}
\textbf{Proof:}
The conditional states $\rho_{B|j}^\mathcal{M}$ can be written as
\begin{equation}
\rho_{B|j}^\mathcal{M}=\sum_i\frac{  p_i q_{ij} }{p'_j} \rho_{B}^i
\end{equation}
being $p'_j=\sum_ip_iq_{ij}$ and $q_{ij}=\textrm{Tr}\left[ M_j  \ket{i} \bra{i}_A \right]$. Therefore, we have
\begin{equation}
\mathcal{J}_d^\mathcal{M} (\rho_c) = \sum_i  \ p'_j \  d \left(\sum_i\frac{  p_i q_{ij} }{p'_j} \rho_{B}^i \ || \ \rho_B \right).
\end{equation}
As the distance $d(\cdot||\cdot)$ fulfills property \ref{PRCC}, the following inequality holds,
\begin{align}
d\left( \sum_{i} \  \frac{p_iq_{ij}}{p'_j}\rho_B^i \ || \ \rho_B \right) \leq \sum_{i} \frac{p_iq_{ij}}{p'_j} \ d\left( \rho_B^i ||\rho_B \right).
\end{align}
Using that $\sum_j M_j = \mathbb{1}$ and $\textrm{Tr}[\ket{i}\bra{i}_A]=1$ it follows that $\sum_j \ q_{ij}= 1$. Thus, by multiplying this last inequality with $p'_j$ and performing the sum over $j$, we obtain
\begin{equation}
\mathcal{J}_d^\mathcal{M}(\rho_c) \leq \sum_i \  p_i \ d(\rho_B^i||\rho_B).
\end{equation}
$\blacksquare$.\\
It should be noticed that we have not imposed the condition of symmetry on $d(\cdot||\cdot)$. Therefore, according to the last proposition, the conditional states $\rho_{B|j}^\mathcal{M}$ must enter in the argument for which the distance is convex.\par
Next, in order to verify that $\mathcal{J}_d$ is a measure of classical-quantum correlations, we will consider an arbitrary state $\rho$ and we will perform a von Neumann measurement $\mathcal{M}=\{M_j\}_{j=1}^m$ on the subsystem $A$. The state of the system after the measurement (with unknown
result, i.e., the observers do not have access to the measurement outcome) is given by
\begin{align}
\label{rhoM}
\rho^\mathcal{M}=\sum_j M_j \otimes \mathbb{1} \rho M_j \otimes \mathbb{1}
\end{align}
where $\rho^\mathcal{M} \in \mathcal{C}_A$ is a classical-quantum state. This state can also be written as,
\begin{align}
\rho^\mathcal{M}=\sum_{j} \ p'_j M_j \otimes\rho_{B|j}^\mathcal{M}\,,
\end{align}
where $\rho_{B|j}^\mathcal{M}$ and $p'_j$ are defined in equations \eqref{rhoBJ} and \eqref{PprimaJ}, respectively. Then, using Eq. \eqref{extra property} and the convexity of $d(\cdot||\cdot)$ with respect to its first input, we have
\begin{align}
\label{demcc}
\mathcal{J}_d(\rho^\mathcal{M})=\sum_{j=1} p'_j \ d\left( \rho_{B|j}^\mathcal{M} || \rho_B \right)=\mathcal{J}^\mathcal{M}_d (\rho).
\end{align}
The state $\rho^\mathcal{M}$ represents a classical-quantum correlated state, in the sense that the correlations are between a classical subsystem $A$ and a quantum subsystem $B$. Since $\rho$ is an arbitrary state, equation \eqref{demcc} allows us to conclude that $\mathcal{J}_d$ genuinely quantifies classical-quantum correlations.
\subsection{Generalized measures of quantum correlations\label{sec:GenMeasurQcorr}}
Having defined generalized measures of total and classical-quantum correlations and keeping in mind Eq.\eqref{eq:genQCorr}, we propose a measure of quantum correlations as follows,
\begin{equation}
\mathcal{Q}_d \left( \rho \right) := \mathcal{T}_d\left( \rho \right) - \mathcal{J}_d\left( \rho \right).
\end{equation}
From previous sections we know that if  $d(\cdot||\cdot)$ satisfies conditions \ref{TP2}, \ref{TP1} and \ref{TP3}-\ref{PRCC} (cf. Sec. \ref{sec:dprop}) then $\mathcal{T}_d$ and $\mathcal{J}_d$ satisfy the necessary conditions \ref{cond 1} - \ref{cond 4} for correlation measures (cf. Sec. \ref{prop measures}). In order for $\mathcal{Q}_d$ to fulfill the necessary criteria (cf. Sec. \ref{prop measures}), we need to impose an extra condition on $d(\cdot||\cdot)$. Thus, we introduce the following proposition:
\begin{description}[font=\bfseries,align=left]
	\item[Proposition V] If $d(\cdot||\cdot)$ has the properties \ref{TP2}, \ref{TP1} and \ref{TP3}-\ref{PRQC} (cf. Sec. \ref{sec:dprop}) then $\mathcal{Q}_d$ meets \ref{cond 1}-\ref{cond 3} and \ref{cond 5} (cf. Sec. \ref{prop measures}).
\end{description}
\textbf{Proof:} The proof of this last proposition relies upon the consistency of the scheme established by the measures $\mathcal{T}_d$, $\mathcal{J}_d$, and $\mathcal{Q}_d$. Thus, we shall prove next that $\mathcal{Q}_d$ is positive-semidefinite.\par
For the classical-quantum state $\rho^\mathcal{M}$ obtained after measurement, given by Eq. (\ref{rhoM}), we can write the following inequality,
\begin{equation}
\label{kii}
\mathcal{T}_d(\rho) \geq \mathcal{T}_d(\rho^\mathcal{M}).
\end{equation}
As $\rho^{\mathcal{M}} \in \mathcal{C}_{A}$, according to the property \ref{PRQC} of $d(\cdot||\cdot)$ it follows that
\begin{align}
\mathcal{T}_d(\rho^\mathcal{M})=\mathcal{J}_d(\rho^\mathcal{M})
\end{align}
or equivalently according with Eq. (\ref{demcc})
\begin{align}
\mathcal{T}_d(\rho^\mathcal{M})=\mathcal{J}_d^{\mathcal{M}}(\rho).
\end{align}
Then, using Eq. (\ref{kii}) we can write
\begin{align}
\mathcal{T}_d(\rho) \geq \mathcal{J}_d^\mathcal{M}(\rho)
\end{align}
for any von Neumann measurement $\mathcal{M}$, in particular for the one that maximizes the classical-quantum correlations. Thus, the condition
\begin{align}
\mathcal{T}_d(\rho) \geq \mathcal{J}_d(\rho)
\end{align}
is satisfied and the positivity of $\mathcal{Q}_d$ has been demonstrated.\\
The proof that  $\mathcal{Q}_d$ satisfies requirement \ref{cond 5} follows from property \ref{PRQC} and Eq. \eqref{demcc}. In addition, due to $\mathcal{T}_d$ and $\mathcal{J}_d$ are  invariant under local unitary transformations and both yield zero when evaluated in product states, it follows that $\mathcal{Q}_d$ also satisfy properties \ref{cond 1} and \ref{cond 2}.$\blacksquare$\\
It should be noticed that the measure of quantum correlations $\mathcal{Q}_d$ reduces to quantum discord in the case $d(\cdot||\cdot)=S(\cdot||\cdot)$, being $S$ the quantum relative entropy.
\subsubsection{Examples}
Previously, we obtained a set of requirements for a quantum distance in order to build well-behaved quantifiers of correlations between subsystems (cf. Sec.\ref{sec:main}). In the case of generalized measures of total and classical-quantum correlations, on the one hand, our results show that distance $d$ should satisfy properties \ref{M1}, \ref{M2} and \ref{TP3}. On the other hand, in the case of quantum correlations measures, distance $d$ should also satisfy property \ref{PRQC} (cf. Sec. \ref{sec:dprop}).\\
Now, besides the well-known case of relative entropy, we shall consider the following particular examples: the \textit{trace distance}, the squared \textit{Bures distance}, the \textit{quantum Jensen-Shannon divergence} (QJSD) and the squared \textit{Hellinger distance}. 
All these distances fulfill the conditions \ref{M1}, \ref{M2} and \ref{TP3} \cite{Nielsen&Chuang,Majtey2005,Spehner2017}. The only requirement not considered in literature is \ref{PRQC}. Therefore we will prove this last property for the cases of trace distance and Quantum Jensen Shannon divergence. The remaining cases can be easily proved using similar arguments to those of trace distance. \par
\textbf{Trace distance:}\\
If $\rho_1$ and $\rho_2$ are two density matrices, then trace distance is defined as follows: \cite{Nielsen&Chuang}:
\begin{equation}
d_{Tr}(\rho_1 || \rho_2) = \frac{1}{2}\textrm{Tr}\left[\sqrt{(\rho_1-\rho_2)^2}\right].
\end{equation}
Let us choose $\rho_1=\sum_i p_i E_i \otimes \rho_i$ and $\rho_2=\sum_j p_j E_j \otimes \rho$, being $E_j=\ket{j}\bra{j}$ and $\{\ket{j}\}$ an orthonormal basis of the subsystem $A$. Then,
\begin{equation}
(\rho_1-\rho_2)^2=\sum_i p_i^2 E_i\otimes (\rho_i-\rho)^2.
\end{equation}
As $\left(\sum_i p_i E_i \otimes \sqrt{(\rho_i-\rho)^2}\right)^2=\sum_i p_i^2 E_i\otimes(\rho_i-\rho)^2$ then
\begin{equation}
\sqrt{(\rho_1-\rho_2)^2}=\sum_i p_i E_i\otimes \sqrt{(\rho_i-\rho)^2}.
\end{equation}
Consequently,
\begin{equation}
d_{Tr}(\rho_1||\rho_2)=\frac{1}{2}\sum_i p_i \textrm{Tr}\left[\sqrt{(\rho_i-\rho)^2}\right]=\sum_i p_i d_{Tr}(\rho_i||\rho).
\end{equation}
\textbf{Quantum Jensen Shannon divergence:}\\
The QJSD is defined as follow
\begin{equation}
d_{js}(\rho_1||\rho_2)=S\left(\frac{\rho_1+\rho_2}{2}\right)-\frac{1}{2}S(\rho_1)-\frac{1}{2}S(\rho_2)
\end{equation}
being $S(\cdot)$ the von Neumann entropy which yields the following result when evaluated in classical-quantum states:
\begin{equation}
S(\rho_1)=-\textrm{Tr}\left[\rho_1\log \rho_1\right]=H(p)+\sum_i p_i S(\rho_i)
\end{equation}
In last equation, $H(p)=-\sum_i p_i \log p_i$ represents the Shannon entropy. Thus,
\begin{align}
&S(\frac{\rho_1 + \rho_2}{2})=H(p)+\sum_i p_i S(\frac{\rho_i + \rho}{2})  \\
&S(\rho_1)=H(p)+\sum_i p_i S(\rho_i) \\
&S(\rho_2)=H(p)+\sum_i p_i S(\rho).
\end{align}
It immediately follows that
\begin{equation}
d_{js}(\rho_1||\rho_2)=\sum_i p_i \left[S\left(\frac{\rho_i + \rho}{2}\right)-\frac{1}{2}S(\rho_i)-\frac{1}{2}S(\rho)\right]=\sum_i p_i d_{js}(\rho_i||\rho)
\end{equation}
\subsubsection{Additional conditions for quantum correlations measures}
In reference \cite{ABC16} Adesso {\it et al.} proposed three additional conditions that should be satisfied for well-behaved measures of quantum correlations. Two of them are related to the behavior of the measure under local quantum operations. The other one states that for a pure state the measure should reduce to an entanglement quantifier. \par
According to ref. \cite{ABC16}, it should be impossible to create quantum correlations by performing operations only on party B of the bipartite system. In this context, we were able to prove the following proposition:
\begin{description}[font=\bfseries,align=left]
	\item[Proposition VI] If $d(\cdot||\cdot)$ satisfy the conditions \ref{TP1}-\ref{PRQC} (cf. Sec. \ref{sec:dprop}) in addition to the following two properties:
	\begin{itemize}
		\item {\textit{Restricted Additivity}},\\
		$d(\rho_1 \otimes \sigma||\rho_2 \otimes \sigma) = d(\rho_1 || \rho_2)$ where $\rho_1$, $\rho_2$ and $\sigma$ belong to $\mathcal{B}_1^+(\mathcal{H})$.
		\item Given $\sigma_{ABE} \in \mathcal{B}_1^+(\mathcal{H}_{ABE})$, $\mathcal{H}=\mathcal{H_A}\otimes\mathcal{H_B}\otimes\mathcal{H}_E$, the quantity
		\begin{equation}
		d(\sigma_{ABE}||\sigma_{AE}\otimes\sigma_B)-d(\sigma_{AB}||\sigma_A\otimes\sigma_B) \label{55}
		\end{equation}
		is invariant under the exchange of the subsystems $B$ and $E$. Here, the matrices $\sigma$ in the arguments of $d(\cdot||\cdot)$ are the reduced matrices of $\sigma_{ABE}$, for example: $\sigma_{AE}=\textrm{Tr}_B [\sigma_{ABE}]$. 
	\end{itemize}
	then $\mathcal{Q}_d(\rho)$ is monotonic non-increasing under local operations on subsystem $B$  (cf. property \ref{cond 4}, Sec. \ref{prop measures}).
\end{description}\par
\textbf{Proof:} In reference \cite{Piani2102} it is demonstrated that quantum discord satisfies property \ref{cond 4}. The proof makes use of the properties of von Neumann entropy. Proposition (VI) can be proved following the same lines of reasoning as in ref. \cite{Piani2102}$\blacksquare$.\par
If some of the assumptions in proposition (VI) are relaxed, it may happen that the monotonically non-increasing behavior of $\mathcal {Q}_d(\rho)$ remain valid. In such a case, a particular analysis should be necessary for the particular choice of the distance $d$.

Finally, it is important to study the conditions that the distance $d$ should satisfy so that $\mathcal {Q}_d$ is reduced to a physically well-behaved entanglement measure in the case of pure states. This is not an easy task and it will be the subject of future research.
\section{Concluding remarks\label{sec:conclusions}}
In this work, we proposed a general approach to the problem of quantifying different types of correlations in bipartite quantum systems through the use of distances between quantum states and processes. Under the assumption that total correlations contained in a bipartite state can be assessed as a sum of classical and quantum contributions, we analyzed the properties that distances must fulfill in order to derive well-behaved quantifiers of correlations. On the one hand, we demonstrated that distances satisfying basic properties, namely non-negativity, identity of indiscernibles, and invariance under unitary transformations, in addition to the requirement of being non--increasing under the action of trace-preserving quantum operations (cf. Sec. \ref{sec:dprop}, properties \ref{TP2}, \ref{TP1}, \ref{TP3}, and  \ref{PRTC}), can be used to build physically well-behaved measures of total correlations. On the other hand, in order to obtain well-behaved measures of classical-quantum correlations, distances must verify, in addition to the basic properties, the requirement of being convex in one of its inputs (cf. Sec. \ref{sec:dprop}, property \ref{PRCC}). Furthermore, in order to obtain well-behaved quantum correlation measures, we showed that distances should also satisfy an additional not too restrictive requirement (cf. Sec. \ref{sec:dprop}, property \ref{PRQC}). We also showed that the proposed measures of quantum correlations are monotonic non-increasing under local operation on subsystem $B$ whenever the distance measure satisfies the restricted additivity and the relation given by Eq.\eqref{55}.

Finally, besides the well-known case of relative entropy, we introduced some additional examples of distance measures, i.e.,  the \textit{trace distance}, the squared \textit{Bures distance}, the \textit{quantum Jensen-Shannon divergence} (QJSD) and the squared \textit{Hellinger distance}, which can be used to build physically well-behaved quantifiers of correlations.
As a result, we showed that it is possible to build quantifiers of different types of correlations within a consistent framework. These quantifiers might be used as additional tools for analyzing different possible scenarios of physical relevance.

\begin{acknowledgements}
	D.B., A.P.M., P.W.L. and T.M.O acknowledge the Argentinian agency SeCyT-UNC and CONICET for financial support. D. B. has a fellowship from CONICET.
\end{acknowledgements}


{}

\end{document}